\documentclass[a4paper,twocolumn,english,showkeys]{revtex4}
\usepackage{pslatex}
\usepackage[T1]{fontenc}
\usepackage[latin1]{inputenc}
\usepackage{amsmath}
\usepackage{graphicx}
\usepackage{amssymb}

\makeatletter

\providecommand{\tabularnewline}{\\}

\usepackage{babel}
\makeatother
\begin{document}

\title{Elements for Response Time Statistics in ERP Transaction Systems}

\author{\textsf{Andreas Mielke}}

\email{am@vm-s.com}

\homepage{http://www.vm-s.com}

\address{\textsf{VM SOLUTIONS GmbH, Heidelberg, Germany}\\
\textsf{~}\\
\textsf{Institut für Theoretische Physik, Univ. Heidelberg, Germany}\\
\textsf{~}\\
\textsf{\today}}

\begin{abstract}
The aim of this work is provide some insight into the response time
statistics of enterprise resource planning systems. We propose a simple
mean-field model for the response-time distribution in such systems.
This model yields a log-normal distribution of response times. We
present data from performance measurements to support the result.
The data show that the response-time distribution of a given transaction
in a given system is generically a log-normal distribution or, in
some situations, a sum of two or more log-normal distributions. Deviations
of the log-normal form can often be traced back to performance problems
in the system. Consequences for the interpretation of response time
data and for service level agreements are discussed.
\end{abstract}


\maketitle

\section{Introduction}

Nearly every large enterprise runs several, sometimes more than 100
ERP (Enterprise Resource Planning) systems like SAP R/3%
\footnote{SAP® and R/3® are registered trademarks of SAP AG, Walldorf, Germany.%
}. These systems are used to electronically support any kind of business
process. The cost of operation of the ERP system landscape in an enterprise
lies typically between 1\% and 4\% of the total revenue of the enterprise,
depending on the type of industry. A good quality of such systems
is important, since essential business processes (production, delivery,
material management, sales and distribution, etc.) depend on these
systems. Lack of quality of an ERP system often produces high cost.

Performance, availability, and security are the most important criteria
for the quality of an ERP transaction system. Whereas availability
and security of such systems are often controlled in detail by service
level agreements, agreed upon performance data are less sophisticated.
Typically one fixes a value for the mean response time, either for
all transactions or for a certain class of transactions. Sometimes
but rarely one finds statements where a certain quantile of the response-time
distribution of all transactions is fixed, e.g. {}``80\% of all transactions
should have a mean response time less than one second''. But it remains
completely unclear whether such a statement is better. What are the
relevant performance measures of an ERP system? The answer to this
question requires a detailed knowledge of the response time statistics
of transactions.

The ERP system landscape in an enterprise is highly dynamical. The
behaviour of users change, new users are working with the system,
systems are consolidated, new systems are introduced, new functionality
is needed, etc. Therefore, a proactive performance management would
be desirable. But unfortunately, a detailed knowledge of relevant
statistical properties of performance data in such systems is still
lacking. The aim of the present work is to gain some insight into
the response time statistics of ERP systems.

\subsection{Relation to performance studies for other systems}

Unfortunately, a realistic theory of the response-time distribution
in transaction systems like R/3 does not exist. To some extent one
may compare such a system with a soft real time system. Trivedi \emph{et
al}. \cite{Trivedi03} recently published a comprehensive review of
models for response-time distributions in real time systems. The analytical
modeling frameworks for such systems are Markov models and stochastic
petri nets. Computing the response-time distribution in these frameworks
is difficult. Closed-form expressions are available only for simple
queuing systems. For networks of queues, a numerical solution using
Markov chains appears to be the only possible method. Recently, AuYeung
\emph{et al}. \cite{Au-Yeung2004} proposed the use of generalized
lambda distributions to obtain an approximation of the response-time
distribution. They applied this approximation to different models
and were able to show that the approximation yields excellent results. 

But the contribution of the queue-time or the wait time to the response
time in an R/3 system is marginal, the relevant contributions to the
response-time distribution of a given transaction are given by the
variation of the DB request time and processing time \cite{Schneider02}.
The design of a transaction system like SAP R/3 tries to avoid long
queue-times and long wait times using several work-processes for dialog
steps and separate update-, enqueue and spool processes. Long queue-times
or long wait times are considered as a performance problem. We will
come back to this point later.

To some extent one can compare performance issues in ERP systems to
web services. Paxson \cite{Paxson94} showed that log-normal distributions
describe the statistics of relevant performance data for simple services
like telnet, nntp, smtp, ftp very well. It has been argued that at
least the body of the actual distributions are close to log-normal,
whereas the (heavy) tails of the distributions are described by power
laws \cite{Crovella95} due to self similarity and fractality \cite{Willinger98,Sole01}.
Compared to the Internet, the structure of an ERP system is less complicated.
There is no self similarity or fractality in ERP systems, and one
should not expect data size distributions with heavy tails. Therefore,
we may expect a log-normal distribution for response times, but no
heavy tails. In fact, heavy tails in the response-time distribution
of a given transaction would be considered as a performance problem
in such systems.

A better system to compare with is a web based shopping system \cite{Arlitt01}.
The structure of a shopping system is similar to that of an ERP system.
It has a multi tier structure with a database server, several application
servers, web servers and clients. The difference to an ERP system
lies in the usage: The main problem in a shopping system is the lack
of control over how many users may arive, whereas in an ERP system
the total number of users is known, their behaviour is predictable,
and the number of concurrent users varies only on time scales long
compared to typical response times. As a consequence we should expect
somewhat broader response-time distributions for the web based shopping
system.

\subsection{Outline of the paper}

From the analogies to Web services or web based shopping systems one
might guess that response time statistics in ERP systems can be described
by log-normal distribributions and that there are no heavy tails.
The aim of this paper is to support that guess. In the next section
we describe in some detail what are the main contributions to the
response time of a transaction in an ERP system. Furthermore we indicate
how these data are measured. We present a simple mean-field model
for the response-time distribution and discuss the validity of that
model.

In Sect. 3 we present measured data for different transactions in
different ERP systems. The data agree well with the assumption of
a log-normal distribution. Deviations of the log-normal form occur
and often indicate a performance problem. 

In Sect. 4 we discuss the results and Sect. 5 contains some conclusions.
The two main consequences are: 

\begin{enumerate}
\item Since deviations of the log-normal distribution indicate a performance
problem, one can try to use response-time distributions to locate
performance problems. Since the number of possible reasons for performance
problems in ERP systems is enormous, we discuss some typical examples
in Sect. 4.
\item The result has some interesting consequences for agreements on the
performance of R/3 systems. The log-normal distribution is skew and
has a tail for longer response times. This means that agreements on
mean values are not suitable in concrete situations. We discuss this
point in some detail in Sect. 5.
\end{enumerate}

\section{Response times in ERP systems}

\subsection{Characteristic times in ERP systems}

Let us start with some remarks on how transactions are executed in
ERP transaction systems. Roughly, the following steps are executed

\begin{enumerate}
\item A user starts a transaction, the request is sent to the dispatcher
processes. 
\item The dispatcher sends the request to a free work process. 
\item The work process executes the transaction. 
\item During the execution, the work process connects to the database to
read or write data.
\item Data are sent back to the user.
\item Data are sent to the update process, which writes data to the database
if necessary.
\end{enumerate}
The steps 3 and 4 (exectution, reading and writing data) may be repeated
several times before step 5 follows. The response time is the time
that is needed for the steps 1 to 5. It consists of the following
components:

\begin{description}
\item [network~time:]The time that is needed to send data from the user
to the system and back. 
\item [wait~time:]The time the dispatcher needs to find a free work process.
\item [load~time:]The time that is needed to load the program into the
work process.
\item [processing~time:]The cpu-time of the work process.
\item [DB~request~time:]The time needed to execute database requests.
\item [enqueue~time:]The time the process has to wait due to queueing.
\item [gui~time:]The time the client needs to build up the graphical user
interface and to show the results.
\end{description}
The amount of data transfered from the user to the system and back
is typically small. Therefore, in modern architectures, the network
time is typically short. This is true for the gui time as well. Todays
clients are powerful and therefore the gui time is short.

Programs are usually kept in the program buffer, so that the load
time should be small.

The hardware architecture of ERP systems typically consists of a database
server and serveral application servers. Logically, the ERP system
is organized as an ensemble of processes. The number of transactions
that can be executed at the same time is determined by the number
of work processes. The number of work processes is typically chosen
to be some multiple of the number of cpus of all application servers.
One usually tries to size the system so that practically no wait time
occurs. 

Different users working at the same time on the system typically use
different transactions or use different data. Therefore it rarely
happens that a transaction has to wait until the data of another transaction
are written to the database. This means that the queue time is typically
short compared to the response time. Furthermore, different modi for
the dispatcher can be chosen to reduce queueing.

This design of a typical ERP system has the consequence that the main
contribution to the response time are the process time and the database
request time.

\subsection{Performance measurements in ERP systems}

Since the performance of an ERP system is an important issue for enterprises
running such systems, ERP systems have builtin performance measurement
tools. For SAP systems, the builtin performance measurement tool is
highly sophisticated and yields detailed performance data for single
transactions or sets of transactions. Therefore we restrict the measurements
presented in this paper to SAP systems (SAP is the market leader with
a market share above 50\%). In SAP systems the set of tools is called
CCMS (Computing Center Management System). The CCMS contains a huge
set of performance data for each transaction: response time, cpu time,
wait time, queue time, network time, gui time, times for direct read
requests, for sequential read requests, for change requests, etc.
The data are available for single transactions but the system agregates
the data so that after some time (an hour, a day, a week, a month)
only sums of such data over a time interval are available. Therefore
we wrote a small program for SAP systems that regularly collects the
data. This program is used to obtain precise performance data of SAP
systems over a long period of time. In total we performed such performance
measurements for more than 250 SAP systems running in many different
enterprises from different industries. The smallest system has 20
users, the largest more than 30,000. The data we show in this paper
are taken from these performance measurements.

Since there are differences between different releases of SAP R/3,
we restrict ourselves to newer releases (the kernel release should
be 4.6D or higher). For older releases, the network time and the gui
time were not included in the response time and therefore one is not
able to obtain a complete picture from response time measurements
in systems with older releases. Nevertheless, since the architecture
of the software has not been changed, we expect that our results apply
to older releases as well.

The clear advantage of the type of measurement we used is that it
allows for performance measurements in a large number of production
systems without having an impact on the performance of these systems.
The disadvantage is that one is limited to the data provided by the
CCMS. 

Although the CCMS allows for detailed performance measurements, it
is often impossible to identify a performance problem proactively.
Tools for an automatic real-time analysis of the data do not exist.
Therefore, in a typical situation, a system administrator uses the
CCMS to find a possible reason for a performance problem that has
been reported by a user. One goal of the present work is to show how
performance problems can be found proactively by a statistical analysis
of the data.

Although we restrict ourselves to measurements in SAP R/3, we expect
the results to be valid for any ERP system. The (hardware and software)
architecture of other products is similar to SAP R/3.

\subsection{Heuristic rules}

In the literature one often finds heuristic rules which help to evalute
measurements of average response times, see e.g. \cite{Schneider02},
p. 134ff. These rules are used by system administrators. Severe violations
of such rules indicate a performance problem, and depending on the
specific rule that is violated one can conclude what has to be done.
Typical rules for response times of dialog steps in SAP systems are
(a simplified version of the table in \cite{Schneider02}, p. 134f.,
Schneider correctly distinguishes between the CPU time and the processing
time):

\begin{center}\begin{tabular}{|l|l|}
\hline 
Performance data&
Time\tabularnewline
\hline
\hline 
Average response time&
$\propto$1 second \tabularnewline
\hline 
Average CPU time&
$\approx$40\% of the average response time\tabularnewline
\hline 
Average wait time&
$<$1\% of the average response time\tabularnewline
\hline 
Average queue time&
$<$1\% of the average response time\tabularnewline
\hline 
Average load time&
$<$10\% of the average response time\tabularnewline
\hline 
Average DB request time&
$\approx$40\% of the average response time\tabularnewline
\hline
\end{tabular}\end{center}

None of these rules represents a clear quantitative statement, but
only an order of magnitude for a given quantity. The first rule, stating
that the average response time for a given transaction for instance
is of the order of a second, means that it may be 300 ms or less,
it may be 3 seconds for more complex transactions, and it depends
clearly on details of the system like the server hardware, storage,
database. Following \cite{Schneider02}, a violation of one of the
heuristic rules may be an indication of a performance problem. 

The main problem is that statements on averages may help if one has
a permanent performance problem. Temporary, casual or periodic performance
problems cannot be found using these rules. On the other hand, the
rules provide an estimate of the different contributions and show
that in a performant system wait time, queue time and load time play
only a minor role.

\subsection{Simple models}

A single transaction in an ERP system runs in a stochastic environment.
Since the system executes many transactions at the same time, resources
are shared between many transactions. In a detailed model, one could
describe a single transaction by a state variable $\xi(t)$ and the
dynamic behaviour of that state variable by a stochastic differential
equation. One can choose $\xi(t)$ so that the transaction starts
at $\xi(t)=0$ and terminates at $\xi(t)=1$. We assume that $\xi(t)$
is a monotonously increasing stochastic Markov process. The probability
distribution $\rho(x,t)$ for that variable $\xi(t)$ can be described
by a Fokker-Planck like equation $\frac{\partial\rho}{\partial t}=L_{FP}\rho$.
The response time $\tau$ for the transaction is the first passage
time for $\xi(t)$ with a starting point 0 and a final point 1. The
response-time distribution $p(\tau)$ is given by $p(\tau)=-\frac{d}{d\tau}\int_{0}^{1}dx\,\rho(x,\tau)$
\cite{Risken}. 

To our knowledge, such a detailed model does not exist. Furthermore,
it would be very difficult to verify such a model, since it is practically
impossible to measure the progress of a transaction without a significant
impact on the performance of the system. To simplify the situation,
let us make some plausible assumptions on an ideal ERP system:

\begin{enumerate}
\item All transactions can be executed at any time without delay. In other
words: necessary data are available and the system has enough resources.
\item Peripheral processes (waiting in the queue, roll-in, etc.) can be
neglected.
\item The state of the system can be described by a set of external parameters
(e.g., the cpu load of the servers). The variation of these parameters
affects all transactions in a similar way.
\item The transaction can be described by a set of internal parameters (like
e.g. the number of data read).
\item The time scale on which the state of the system varies is long compared
to typical response times. 
\end{enumerate}
The first and the second condition are idealized consequences of the
heuristic rules and reflect the design of a transaction system like
SAP R/3. In SAP systems, one tries to realize the second condition
using a sufficient number of work processes. 

The last condition simply means that a given transaction running in
the system {}``sees'' a static environment. Changes of that evironment
are slow. Thus, a theory of response times becomes a static theory:
The response time for a single dialog step depends on quasi static
external parameters describing the system and on internal parameters
describing the transaction. 

With these assumptions we can describe the response time as a function
of the internal and external parameters, $\tau=\tau(\mathbf{a})$.
$\mathbf{a}=(a_{1},\ldots,a_{n})$ denotes the parameters. If one
knows the distribution $P(\mathbf{a})$ of the parameters, the response-time
distribution can be written as \begin{equation}
p(\tau)=\int dP(\mathbf{a})\delta(\tau-\tau(\mathbf{a}))\label{eq:meanfield}\end{equation}
where $\delta(.)$ denotes the Dirac delta distribution. Such a description
can be understood as a mean field approximation of the detailed dynamic
model sketched above. 

In a next step we will make a scaling assumption for $\tau(\mathbf{a})$.
Suppose that one of the parameters is multiplied by a factor $\lambda.$Then
we assume that\begin{equation}
\tau(a_{0}\ldots a_{i-1},\lambda a_{i},a_{i+1}\ldots a_{n})=\lambda^{s_{i}}\tau(\mathbf{a})\label{eq:scaling}\end{equation}
with some exponent $s_{i}$. Let us motivate this assumption by some
simple examples. Suppose that the resource demands such as CPU time
or disk operations on the system are doubled, which means $\lambda=2$
for one of the parameters. Then we would expect that the response
time doubles as well so that the corresponding exponent $s$ would
be 1. Suppose that the amount of data read from the database is doubled.
Then, depending on what is actually done with the data, we would expect
that the DB request time and the processing time become at least twice
as long, and therefore the response time becomes at least twice as
long. The corresponding exponent is therefore 1 or larger. 

This scaling assumption is the main ingredient to our mean-field theory.
It is clear that some of the smaller contributions to the response
time like the wait time and the load time do not depend on all the
parameters mentioned above. Therefore it is crucial that these contributions
are really small and can be neglected. Otherwise, (\ref{eq:scaling})
would not be true.

A direct consequence of (\ref{eq:scaling}) is \begin{equation}
\tau(\mathbf{a})=\tau_{0}\prod_{i}\left(\frac{a_{i}}{a_{i0}}\right)^{s_{i}}\label{eq:tau_ansatz}\end{equation}
and therefore\begin{equation}
\ln(\frac{\tau(\mathbf{a})}{\tau_{0}})=\sum_{i}s_{i}\ln(\frac{a_{i}}{a_{i0}})\label{eq:log_tau}\end{equation}
If we now assume that correlations between the parameters $a_{i}$
are unimportant, the right hand side of (\ref{eq:log_tau}) is a sum
of many independent stochastic variables and the central limit theorem
applies. As a consequence, the distribution of $\ln\tau$ is a normal
distribution and \begin{equation}
p(\tau)d\tau=\frac{1}{\tau\sigma\sqrt{2\pi}}\exp\left(-\frac{1}{2\sigma^{2}}(\ln(\frac{\tau}{\tau_{m}}))^{2}\right)d\tau\label{eq:lognormal}\end{equation}

Here, $\tau_{m}$ is the median, the mean value is $\tau_{av}=\tau_{m}\exp(\sigma/2)$
and the variance is $\tau_{av}^{2}(\exp(\sigma^{2})-1)$. 

To sumarize, variations in the internal and external parameters can
be described by multiplicative noise. Generically, multiplicative
effects of noise yield a log-normal distribution.

\subsection{Deviations from the ideal system}

The assumption of an ideal system, as made above, are often violated.
Let us comment on some of the more important violations.

\begin{itemize}
\item The scaling hypotheses for $\tau$ in (\ref{eq:scaling}) excludes
contributions to $\tau$ like the wait time, the gui time, the load
time. Such contributions would give an additive contribution to $\tau$.
We argued above that in a real system these contributions are small
so that they lead to small corrections to (\ref{eq:lognormal}). A
large value of one of these contributions would be considered as a
performance problem.
\item In a real ERP system a single contribution to the sum in (\ref{eq:log_tau})
can be much larger than any other contribution. This happens, e.g.,
if a bottle neck exists, which is clearly a performance problem. In
fact, the heuristic rules mentioned above suggest that such a domination
should be considered as a performance problem.
\item In a real system the parameters $a_{i}$ will be correlated. Correlations
can be minimized by searching for a suitable set of parameters. In
fact, we have some freedom in the coice of the parameter set. We only
have to make sure that a given set describes the state of the system
and of the transaction sufficiently well. But we cannot expect that
correlations vanish entirely. We can expect that weak correlations
alter the log-normal distribution not too much. We come back to the
discussion of correlations later.
\end{itemize}
Any deviation from an ideal system can yield a performance problem.
And generically, such a deviation causes a violation of our scaling
hypotheses or introduces strong correlations. Therefore, a deviation
from an ideal system generically yields a deviation from the log-normal
distribution for the response times. We will present some examples
below.

\section{Measured response-time distributions}

\subsection{Systems and transactions}

We measured performance data for transactions in 258 systems. In this
section we show representative examples for the response-time distributions
for some transactions in three different systems. Some characteristic
data for these systems are:

\begin{center}\begin{tabular}{|c|r|c|}
\hline 
&
\# user&
\# dialog steps\tabularnewline
\hline
\hline 
System 1&
620&
21,000 steps/hour\tabularnewline
\hline 
System 2&
3,370&
63,000 steps/hour\tabularnewline
\hline 
System 3&
1,140&
124,000 steps/hour\tabularnewline
\hline
\end{tabular}\end{center}

The number of users shown here is the average number of users that
are active during one month. From the point of view of performance,
users in an R/3 system are often classified as occasional users (less
than 100 dialog steps per day), active users (between 100 and 1,000
dialog steps per day), and power users (more than 1,000 dialog steps
per day). System 2 has a high portion of occasional users ($\approx$
68\%), system 3 has many power users ($\approx$ 11\%). The usage
intensity in these three systems is quite different. We performed
response time measurements as described above over a representative
time interval of at least two months.

The number of possible transactions in an SAP system is very large.
For a given system, many transactions are rarely used and it is difficult
to obtain good data. For the three systems we show response time data
for the three transactions VA01 (Create Sales Order), VA02 (Change
Sales Order) and SESSION\_MANAGER. These transactions are among the
most strongly used transactions in the systems 1-3.

\subsection{Examples with a good performance}

Let us first show results of performance measurements in situations
where the performance was good.

\begin{figure}[h]
\begin{center}\includegraphics[%
  width=0.90\columnwidth,
  keepaspectratio]{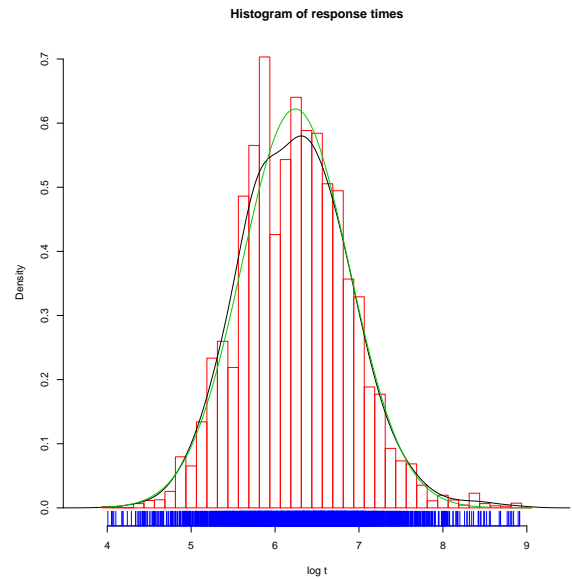}\end{center}

\caption{Transaction VA01, \label{cap:-VA01,-System-1}System 1: The figure
shows the data points (logarithm of the response time in units of
1 ms, blue), a histogram (red), the density distribution (black) and
the normal distribution with the same mean and variance. $\ln\, t=6$
corresponds to a response time $t\approx400$ms, $\ln\, t=8$ corresponds
to $t\approx3000$ms. The basis are 324,000 data points.}
\end{figure}

Figure \ref{cap:-VA01,-System-1} shows results for VA01 in system
1. The distribution agrees quite well with a log-normal distribution,
with a small deviation for $\tau\approx\tau_{m}$. The log-normal
distribution has been calculated from the mean and variance of the
actual data, no parameters have been adjusted. The data for VA02 and
SESSION\_MANAGER in system 1 and for SESSION\_MANAGER in system 2
show a similar behaviour. Figs. \ref{cap:-VA01,-System-1} to \ref{cap:SESSION_MANAGER,-System-2:}
show an impressive agreement of the actual data with a log-normal
distribution. 

\begin{figure}[h]
\begin{center}\includegraphics[%
  width=0.90\columnwidth,
  keepaspectratio]{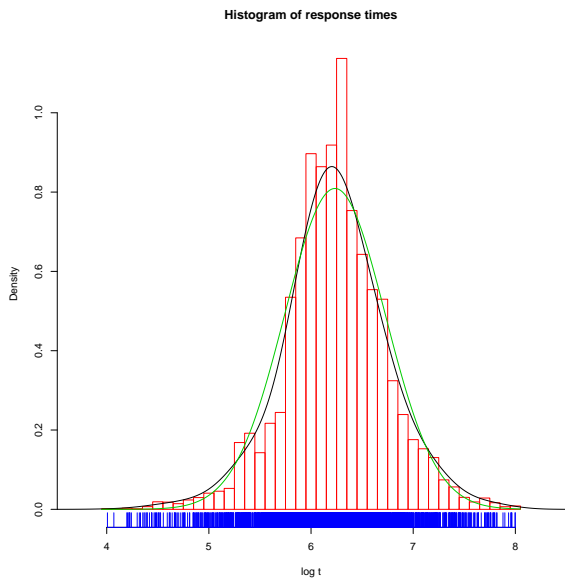}\end{center}

\caption{Transaction VA02, System 1: For details see Fig. \ref{cap:-VA01,-System-1}.
Basis: 419,000 data points.}
\end{figure}

\begin{figure}[h]
\begin{center}\includegraphics[%
  width=0.90\columnwidth,
  keepaspectratio]{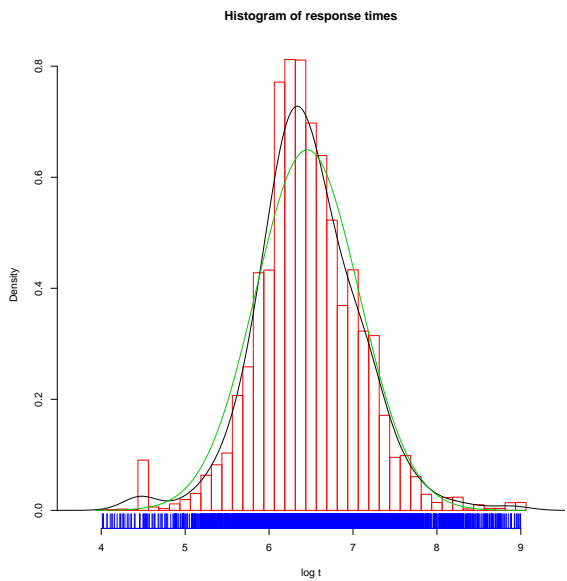}\end{center}

\caption{SESSION\_MANAGER, System 1: For details see Fig. \ref{cap:-VA01,-System-1}.
The basis are 296,000 data points.}
\end{figure}

\begin{figure}[h]
\begin{center}\includegraphics[%
  width=0.90\columnwidth,
  keepaspectratio]{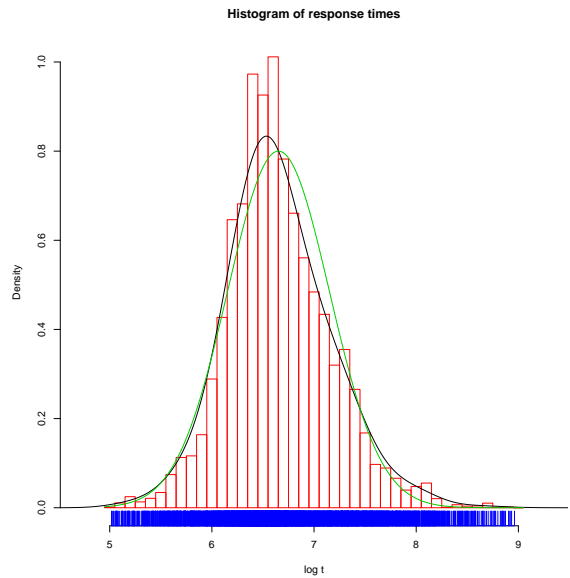}\end{center}

\caption{SESSION\_MANAGER, \label{cap:SESSION_MANAGER,-System-2:}System 2:
For details see Fig. \ref{cap:-VA01,-System-1}. Basis: 1,095,000
data points.}
\end{figure}

The agreement with the log-normal form does not only hold for the
body of the response-time distributions, but, as far as we can judge
from the data, also for the tails. The data show no hint of heavy
tails.

\subsection{Examples with a performance problem}

It is interesting to study some examples where the response-time distribution
differs from the log-normal form.

\begin{figure}[h]
\begin{center}\includegraphics[%
  width=0.90\columnwidth,
  keepaspectratio]{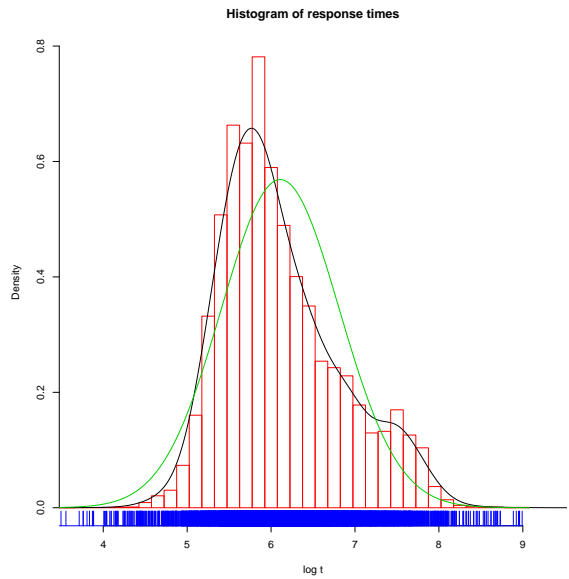}\end{center}

\caption{\label{cap:VA01,-System-2:}VA01, System 2: For details see Fig.
\ref{cap:-VA01,-System-1}. Basis are 143,000 data points.}
\end{figure}

\begin{figure}[h]
\begin{center}\includegraphics[%
  width=0.90\columnwidth,
  keepaspectratio]{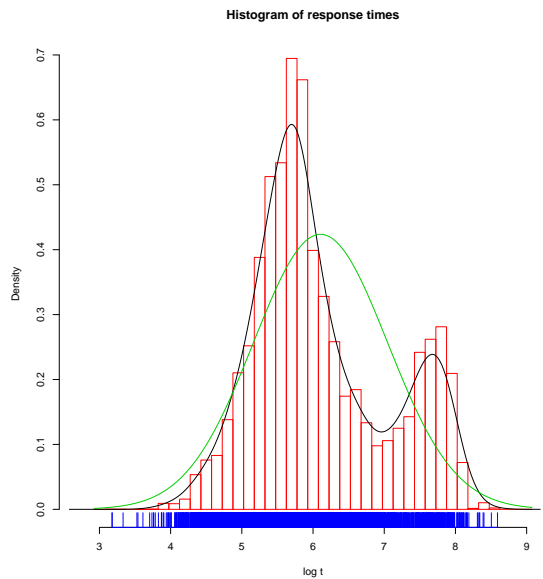}\end{center}

\caption{\label{cap:VA02,-System-3:}VA02, System 3: For details see Fig.
\ref{cap:-VA01,-System-1}. Basis: 281,000 data points.}
\end{figure}

The next two examples, VA01 in system 2 (Fig. \ref{cap:VA01,-System-2:})
and VA02 in system 3 (Fig. \ref{cap:VA02,-System-3:}) differ from
a log-normal distribution. Esp. Fig. \ref{cap:VA02,-System-3:} shows
a clear deviation from the log-normal distribution, the density curve
has a bimodal form. This happens for instance if the transaction runs
in two (or more) distinct situations with clearly different sets of
parameters so that the distribution becomes a sum of two (or more)
log-normal distributions. A similar, but less clear deviation can
be observed in Fig. \ref{cap:VA01,-System-2:}. For both cases one
can show that the set of original data can be decomposed into two
subsets, belonging to different parameter regimes. In both cases,
the system was afflicted with a bottleneck during the period of measurement.
The deviation from the log-normal form indicates a performance problem
in the system. 

A bimodal density curve as shown in the two examples is an indication
of a performace problem. But the form of the density curve is not
sufficient to find out, what kind of performance problem it is and
how it can be cured. As mentioned above, we use the CCMS of the SAP
system to collect many performance data, not only the response time.
In a case like VA01 in system 2 (Fig. \ref{cap:VA01,-System-2:})
one has to analyse these data: Does the problem occur generically,
periodically or occasionally, does it occur for all users or for a
special class of users, which contribution to the response time is
responsible for the problem, etc. In the example VA01 in system 2
it turned out, that the problem occured for a special class of users,
it could finally be traced back to a long time for sequential reads
on a certain table of the database. After having solved the problem,
the density curve had a log-normal form.

There are many different situations which yield a bimodal distribution
curve. A simple example is a real bimodal situation, where the system
runs on several, non-equivalent application servers. Another example
is a system where, from time to time, periodically or casually, a
bottleneck occurs. The actual reason for such a behaviour cannot be
deduced from the response time data alone, but one needs a more complete
set of performance data. 

From a more general point of view the problem is that the response
time for a given transaction as a function of its starting time may
show correlations. Correlations occur because some of the external
parameters (like CPU load or database load) vary in time, but on a
time scale that is long compared to a typical response time. Correlations
may be periodic in time because of a periodic usage of the system.
Deviations from a log-normal distribution may occur due to such a
periodic usage, as was the case in Fig. \ref{cap:VA02,-System-3:}:
This is a typical case for a periodic performance problem in a system.
A similar form of the distribution can be observed if only a special
class of users is affected by a performance problem, if the performance
problem occurs due to a bottle neck in only one of several application
servers, or in similar situations. In all of these cases the response
time data are correlated.

Up to now we have shown results for transactions that are often used.
But the overall result remains true even if a transaction is rarely
used. As an example we finally show results for a proprietary transaction
in system 1. The response-time distribution (Fig. \ref{cap:ZT371,-System-1})
is again well described by a log-normal distribution; due to the smaller
number of data points the fluctuations are larger. Furthermore the
typical response time is longer, the median is $\tau_{m}\approx$3000
ms. 

\begin{figure}[h]
\begin{center}\includegraphics[%
  width=0.90\columnwidth,
  keepaspectratio]{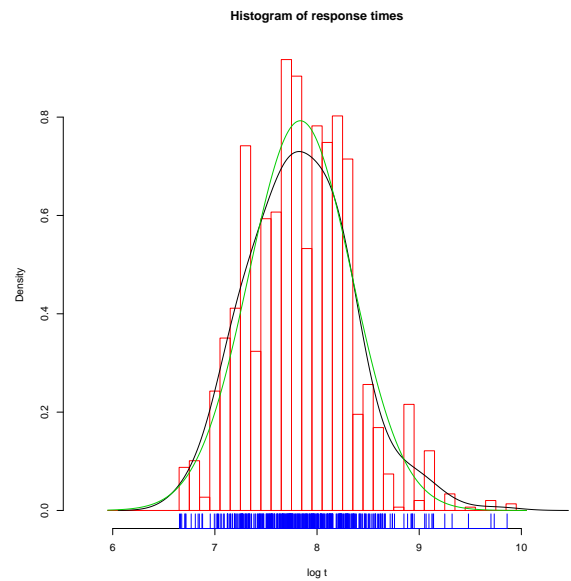}\end{center}

\caption{Transaction ZXXX, \label{cap:ZT371,-System-1}System 1: For details
see Fig. \ref{cap:-VA01,-System-1}. Basis: 1,722 data points. }
\end{figure}

\subsection{General discussion of the log-normal form.}

The above distributions of $\ln\tau$ are representative examples.
Similar calculations can be done for many other transactions and for
other systems. We made of performance measuremens in many SAP R/3
systems to identify and solve performance problems. For each case
one can calculate how much the actual distribution of $\ln\tau$ differs
from a normal distribution. 

\begin{figure}[h]
\begin{center}\includegraphics[%
  width=0.90\columnwidth,
  keepaspectratio]{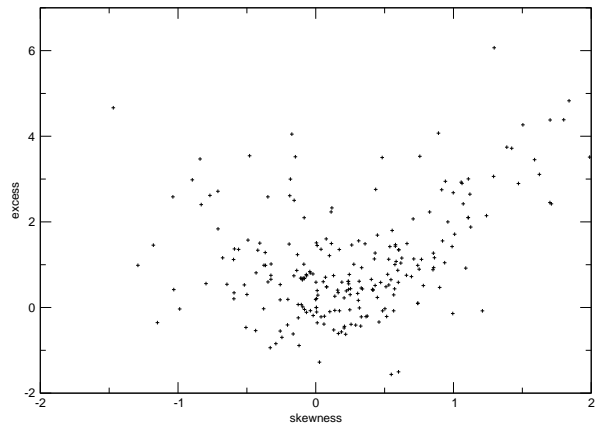}\end{center}

\caption{skewness vs. excess for 232 response-time distributions from 21 different
SAP R/3 systems. \label{cap:restat}}
\end{figure}

The difference of a distribution from the normal form can be described
quantitatively by the higher cumulants $\kappa_{n}$, $n\geq3$. Using
the characteristic function \begin{equation}
\phi(t)=\int dx\, p(x)\exp(itx)\end{equation}
of a distribution function $p(x)$ one defines the cumulants as the
coefficients of the series\begin{equation}
\ln\phi(t)=\sum_{n=0}^{\infty}\kappa_{n}\frac{(it)^{n}}{n!}\end{equation}
 (see \cite{Abramowitz}, number 26.1.12). The third and all higher
cumulants vanish for a normal distribution, since the Fourier transform
of a Gaussian is a Gaussian. Instead of the cumulants we calculate
normalized cumulants $c_{n}=\kappa_{n}/\kappa_{2}^{n/2}$. $c_{3}$
is often called skewness, $c_{4}$ is the excess (or excess kurtosis)
of the distribution. In Fig. \ref{cap:restat} we show a $c_{3}$
vs. $c_{4}$ plot for measured distributions of $x=\ln\tau$ for various
systems and transactions. For a log-normal distribution of $\tau$,
i.e. a normal distribution of $x$, one would have $c_{3}=c_{4}=0$.
For a majority of distributions the skewness is small ($|c_{3}|\lesssim1)$
and the excess is small as well ($0\lesssim c_{4}\lesssim1.5)$. On
the other hand, the figure shows that larger deviations from the normal
form occur. A negative value of $c_{4}$ often occurs in a situation
with two or more maxima (e.g. in Fig \ref{cap:VA02,-System-3:}).
Furthermore, there is a clear tendency towards a positive skewness,
which means that the statistical weight in the tail of the response-time
distribution at long response times is even under-estimated by a log-normal
distribution. In most of the cases shown in Fig. \ref{cap:restat}
a strong deviation from the log-normal form can be traced back to
a performance problem in the system, where one of the heuristic rules
mentioned above is violated. 

Although the data in Fig. \ref{cap:restat} indicate that for a large
subset the distribution function is close to a log-normal form, deviations
occur. We may ask, whether a different form of the distribution function
would give a better fit for these cases. A good candidate is certainly
the generalized lambda distribution (GLD) discussed in \cite{Karian2000,Au-Yeung2004}.
The GLD is a four parameter distribution. It has the ability to assume
a wide variety of shapes. Whereas we used the first and the second
moment to obtain the log-normal distribution shown in the previous
examples, Au-Yeung et al. \cite{Au-Yeung2004} take the skewness $c_{3}$
and the excess $c_{4}$ to determine the two additional parameters.
A point that has to be taken into account is that in Fig. \ref{cap:restat}
we plotted the skewness $c_{3}$ and the excess $c_{4}$ of the distribution
of $\ln\tau$, whereas for the calculation of the GLD as described
in \cite{Au-Yeung2004} we need the skewness $c_{3}$ and the excess
$c_{4}$ of the distribution of $\tau$. 

Let us briefly mention the main conclusion of our efforts using the
GLD. As described in \cite{Au-Yeung2004} this distribution yields
quite good results in situations with heavy tails, where the log-normal
distribution is not applicable. But a heavy tail in the distribution
would be considered as a performance problem and should not occur
in a performant ERP system. On the other hand, the problem is that
the GLD never describes a bimodal density curve. This means that the
examples in section 3.3 cannot be described by a GLD. The same is
true for most of the distribution functions represented by the points
in Fig. \ref{cap:restat}. In most cases, where the probability distribution
is not close to a log-normal form, the GLD fit of the probability
distribution functions looks equally poor.

\section{Discussion of the model}

The simple model we introduced is a mean-field model. It is based
on a set of assumptions expected to hold for an ideal system. The
main ingredient is that $\ln\tau$ can be written as a sum of many
contributions (\ref{eq:log_tau}) and that correlations between the
different contributions can be neglected.

The time for the central processes (reading, writing, processing data)
yields the main contribution to the response time. Therefore one would
naively expect that the corresponding time distributions are connected
to the response-time distribution in a similar manner. A first idea
could be to study the DB request time and the CPU time independently
and to neglect any other contribution. But such an idea is misleading:
Even if one could write the CPU time and the DB request time in a
similar form as the response time in (\ref{eq:log_tau}), the number
of terms in the sum would be smaller and therefore the central limit
theorem cannot be applied as well. To illustrate that such an idea
is misleading, let us show the CPU time distribution and the DB request
time distribution for the transaction VA01 in system 1. Whereas the
response-time distribution in Fig. \ref{cap:-VA01,-System-1} is close
to a log-normal form, the CPU time distribution (Fig. \ref{cap:-VA01,-System-1cpu})
and the DB request time distribution (Fig. \ref{cap:-VA01,-System-1db})
look quite different. The CPU time distribution shows clear deviations
from a log-normal form. On the logarithmic time scale the distribution
is skew. The DB request time distribution shows two maxima at 90 ms
and at 1000 ms. The flanks are much steeper than for a log-normal
distribution. 

\begin{figure}[h]
\begin{center}\includegraphics[%
  width=0.90\columnwidth,
  keepaspectratio]{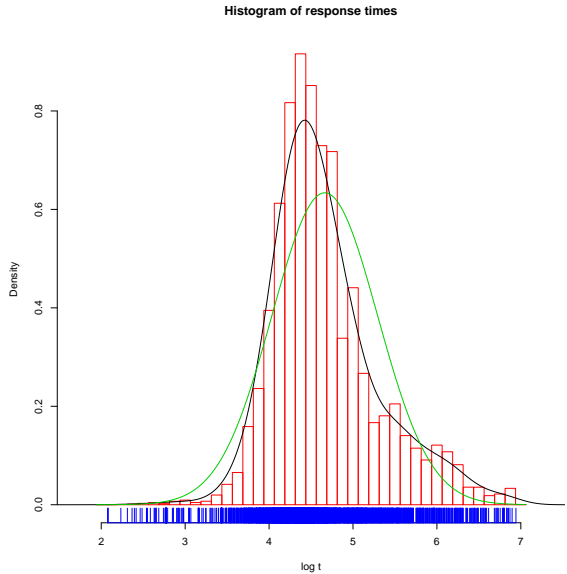}\end{center}

\caption{Transaction VA01, \label{cap:-VA01,-System-1cpu}System 1: The figure
shows the data points (logarithm of the response time in units of
1 ms, blue), a histogram (red), the density distribution (black) for
the CPU time. The basis are 324,000 data points similar to Fig. \ref{cap:cor-VA01,-System-1}.}
\end{figure}

\begin{figure}[h]
\begin{center}\includegraphics[%
  width=0.90\columnwidth,
  keepaspectratio]{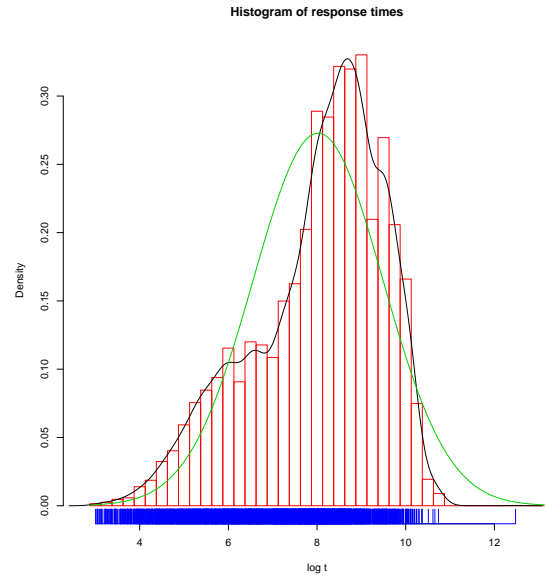}\end{center}

\caption{Transaction VA01, \label{cap:-VA01,-System-1db}System 1: The figure
shows the data points (logarithm of the response time in units of
1 ms, blue), a histogram (red), the density distribution (black) for
the DB request times. The basis are 324,000 data points, similar to
Fig. \ref{cap:cor-VA01,-System-1}}
\end{figure}

Furthermore, the DB request time and the CPU time are correlated.
The reason is that a transaction reads a certain amount of data from
the database, performs operations on these data and eventually writes
data to the database. Both, the CPU time (perfoming operations) and
the DB request time (reading and writing data) depend on the same
data. 

\begin{figure}[h]
\begin{center}\includegraphics[%
  width=0.90\columnwidth,
  keepaspectratio]{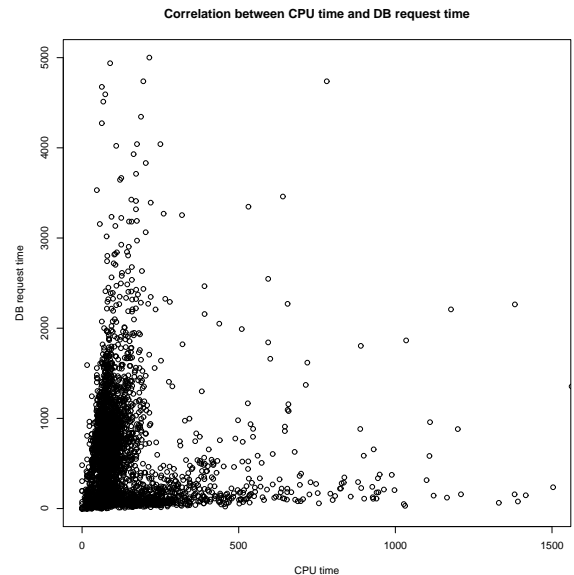}\end{center}

\caption{Transaction VA01, \label{cap:cor-VA01,-System-1}System 1: The plot
shows CPU time vs. DB request time. }
\end{figure}

The plot of the CPU time vs. DB request time for VA01 in system 1,
Fig. \ref{cap:cor-VA01,-System-1}, shows two cleary distinct regions,
one where the CPU time is longer, the other where the DB request time
is longer. Clearly the two maxima in Fig. \ref{cap:-VA01,-System-1db}
correspond to the two regions in Fig. \ref{cap:cor-VA01,-System-1}. 

In different systems or for different transactions the form of the
distribution of the CPU time or the DB request time differ from the
examples shown here. In most cases one learns more about the correlations
in a suitable visulation of the data, as shown above, than using a
sophisticated statistical method. There seems to be no universal form
for the distribution of the CPU time or the DB request time. In contrast,
our analysis indicates that the response-time distribution has a universal
form, as long as we restrict ourselves to a {}``normal'' situation
without performance problems or other specialities. 

Due to the correlations between the CPU time and the DB request time
the joint distribution of these two quantities is not simply the product
of the two distributions shown in Figs. \ref{cap:-VA01,-System-1cpu}
and \ref{cap:-VA01,-System-1db}. This means that the knowledge of
these two distributions does not help if one wants to calculate the
response-time distribution. In other words: In a complete theory correlations
between CPU time and DB request time will play an essential role.

\section{Conclusions and Outlook}

\subsection{General remarks}

The main result of this work is that the typical response-time distribution
of transactions in an R/3 system has a log-normal form. This observation
has several consequences for the interpretation of response time patterns
in R/3 systems. The main point is that a log-normal distribution is
skew and has a long tail for longer response times. As a consequence,
the mean response time alone is not a good performance criterion.
If the shape parameter $\sigma$ of the distribution, i.e., the variance
of the normal distribution of $\ln\tau$, is large, long response
times (twice or three times the mean response time) occur quite often.
Without additional information about the distribution, e.g. the variance,
one is not able to say whether or not the performance of the system
is sufficiently good. 

But the situation is even worse: Often one does not look at the response-time
distribution of a single transaction, but at the response-time distribution
of all dialog steps of a given group of transactions or reports, e.g.,
of all transactions and reports of a given module. If one observes
for such a group of transactions a larger portion with longer response
times, one often argues that this is due to batch-like reports running
in dialog mode. This may of course be true, but another explanation
may be the long tail of the log-normal distribution. This is a main
difference: Whereas a user often knows that a certain report has a
long run-time, he does not expect long response times for typical
dialog transactions. Since the expectation of the user is different,
the user satisfaction will be different as well. 

In Sect. 2 we mentioned heuristic rules for the interpretation of
response times. Although these rules help a lot in performance optimisation
of R/3 systems, they have a severe drawback. This becomes clear when
one looks at a distribution like the one shown in Fig. \ref{cap:VA02,-System-3:}.
The second, smaller peak may be a hint to a performance problem. The
maximum of this peak occurs at $\tau\approx$2.5 s. But the weight
of this peak is less than 10\%, so that it will never be observed
in averages. On the other hand, although there are correlations among
the different contributions to the response time, the heuristic rules
hold only for averages. 

Deviations of the response-time distribution from a log-normal form
may occur. A measure for the deviation is the skewness and the excess
of the distribution of $\ln\tau$. Even if the mean response time
is sufficiently short and if the variance is not too large, a large
skewness or a large excess indicate that a large portion of dialog
steps has long response times. Such deviations indicate a performance
problem in the system. In other words: A plot like the one in Fig.
\ref{cap:restat} for a set of transactions in a given system can
be used to identify problems in that system which cannot be seen regarding
only averages and variances. Often, such performance problems are
not even reported by users, esp. if it concerns an occasional user
who considers the occasionally long response times as 'normal'.

\subsection{Service level agreements}

Another important aspect concerns service level agreements. It is
clear that a simple agreement about the mean response time of all
transactions is not suitable. On the other hand, simply due to practical
restrictions, agreements on performance must be simple and it must
be easy to verify them. One needs a possibility to measure the quantities
that one uses in such an agreement. 

We already mentioned that most ERP system contain builtin tools to
measure response times. Builtin reports yield only averages for the
response times of transactions, but it is a simple task to write a
small program that uses the builtin functionality to calculate other
statistical information as well. 

From the above discussion one would suggest the following rules:

\begin{itemize}
\item Agreements on specific transactions are better than global statements
on averages over a large group of transactions, since the behaviour
of different transactions is different and bottlenecks affect different
transactions in a different way. But if the number of different transactions
is too large, one should restrict agreements to a small set of transactions,
e.g., the ten most important transactions in a system.
\item In addition to a mean response time, one needs a second parameter
to control the variance of the distribution. But this is only suitable
in a situation without a bottleneck, where the form of the response-time
distribution is close to log-normal. Generically, an agreement that
states that a certain portion of all dialog steps for certain transactions
should have a sufficiently small response time (e.g. {}``80\% of
all dialog steps of a given class of transactions should have a response
time less than one second'') is much better than an agreement on
averages, since it allows some control on the tail of the response-time
distribution. 
\end{itemize}
Which class of transactions is actually chosen and what quantile is
suitable depends on the actual system for which a service level agreement
is needed.

\subsection{Future work}

There are three main directions for future work on peformance of ERP
systems:

\begin{enumerate}
\item A much more detailed analysis of the data is needed to better understand
correlations between the different contributions to the response time. 
\item A microscopic model is needed to better understand the deviations
from log-normal form. Such a model should also be able to include
and explain correlations between the different parameters in the system. 
\item There are many reasons why the complexity of the ERP system landscape
in large enterprises is growing: more business processes are supported
by such systems, due to legal reasons additional functionality is
needed, etc. Therefore new types of such systems with special functionality
are needed. Examples are business warehouse systems, customer relationship
management systems, supply chain management systems. It is not clear
whether or how the results presented in this paper can be generalized
to such new systems.
\end{enumerate}
The general goal behind these steps is to obtain a detailed understanding
of the response time statistics of ERP systems.

\end{document}